# The rate coefficients of unimolecular reactions in the systems with power-law distributions

Yin Cangtao, Guo Ran and Du Jiulin*

*Department of Physics, School of Science, Tianjin University, Tianjin 300072, China*

**Abstract:** The rate coefficient formulae of unimolecular reactions are generalized to the systems with the power-law distributions based on nonextensive statistics, and the power-law rate coefficients are derived in the high and low pressure limits, respectively. The numerical analyses are made of the rate coefficients as functions of the $\nu$-parameter, the threshold energy, the temperature and the number of degrees of freedom. We show that the new rate coefficients depend strongly on the $\nu$-parameter different from one (thus from a Boltzmann-Gibbs distribution). Two unimolecular reactions, $CH_3CO \rightarrow CH_3 + CO$ and $CH_3NC \rightarrow CH_3CN$, are taken as application examples to calculate their power-law rate coefficients, which obtained with the $\nu$-parameters slightly different from one can be exactly in agreement with all the experimental studies on these two reactions in the given temperature ranges.

**Keywords**：Reaction rate theory, power-law distributions, unimolecular reaction, nonextensive statistics

## 1. Introduction

Unimolecular reactions are in principle the simplest types of chemical reaction that can occur in the gas phase since the reaction formally involves only one molecule. The theory of unimolecular reactions in the gas phase is a classical topic in physical chemistry [1]. When a molecule is supplied with an amount of energy that exceeds some threshold energy, the unimolecular reaction can take place, that is, dissociation or isomerization. In more advanced treatments of the unimolecular reactions, the mechanism, in the general case of the presence of a foreign gas, has the following scheme [2]:

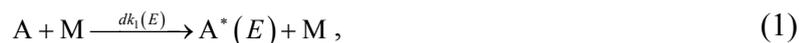

$$A + M \xrightarrow{\ dk_1(E)\ } A^*(E) + M \,, \tag{1}$$

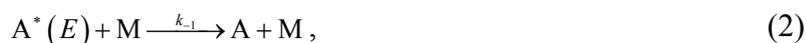

$$A^*(E) + M \xrightarrow{\ k_{-1}\ } A + M \,, \tag{2}$$

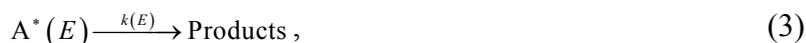

$$A^*(E) \xrightarrow{\ k(E)\ } Products \,, \tag{3}$$

where M is an (inert) buffer gas, or any molecule that does not react with the molecule A, which could be A itself. $A^*$ is a highly vibrationally excited molecule A.

$A^*$ is created in an activation step in reaction (1), which is a thermal activation, due to inelastic collisions, where translational energy is converted to vibrational energy. $dk_1(E)$ is the rate coefficient for activation of the molecule to an energy of the range $[E, E+dE]$. Reaction (2) means that $A^*$ can be de-energized with a rate coefficient $k_{-1}$. Note that $k_{-1}$ is assumed to be independent of the energy. After (or during) the creation of $A^*$, intramolecular vibrational energy redistribution (IVR) will take place, and thus the vibrational energy will redistribute and spread to all vibrational degrees of freedom in the molecule. In reaction (3), the reaction rate coefficient $k(E)$ is assumed a constant reaction probability per unit time at the internal energy $E$. For the activated molecule, the dynamical equation is

$$\frac{dA^*(E)}{dt} = dk_1(E)[A][M] - k_{-1}\left[A^*(E)\right][M] - k(E)\left[A^*(E)\right], \qquad (4)$$

where $[\ldots]$ denotes the concentration. There exists a steady state approximation, and then we have

$$A^*(E) = \frac{dk_1(E)[A][M]}{k_{-1}[M] + k(E)}. \qquad (5)$$

The rate of the reaction is

$$\upsilon = k(E)A^*(E) = \frac{k(E)dk_1(E)[A][M]}{k_{-1}[M] + k(E)}. \qquad (6)$$

Note that the rate, in general, is a function of $[M]$.

In the high pressure limit, the rate is given by a first order expression in $[A]$, and thus from Eq.(6) we have the unimolecular reaction rate coefficient at the energy $E$ [2],

$$k_{uni}^{\infty}(E) = \lim_{[M]\to\infty}\frac{\upsilon}{[A]} = \frac{k(E)dk_1(E)}{k_{-1}}. \qquad (7)$$

At the high pressure, the rates of the energization and de-energization are relatively fast, which thus can be treated as fast pre-equilibrium steps. The rate determination step is the transformation of $A^*$ into products.

Except for high vacuum systems, where isolated reactions may occur, the internal energy is always not fixed. A transition from microscopic to macroscopic description must be taken into consideration. If the rate of the reaction (3) is small as compared with the rates of the activation (1) and the deactivation (2), from Eq. (5) we have $dk_1(E)[A][M] = k_{-1}[A^*(E)][M]$. If we use $P(E)$ to denote the energy distribution function, then the $P(E)dE$ is the probability of finding an A molecule in the range of energy $[E, E+dE]$, ie, $P(E)dE=[A(E)]/ [A]$. Because $A^*$ is A at the high energy, when $E$ is large we have $P(E)dE=[A^*(E)]/ [A]$ and [2] we get

$$\frac{dk_1(E)}{k_{-1}} = P(E)dE, \qquad (8)$$

The rate coefficient counting all the energy (the so-called the thermal rate coefficient), $k_{uni}^{\infty}$, is the integration of Eq.(7) over the energy,

$$k_{uni}^{\infty} = \int_{E_0}^{\infty} k(E) P(E) dE , \qquad (9)$$

where $E_0$ is the threshold energy, and $P(E)$ is BG distribution $P(E)=Q^{-1} N(E)$ $\exp(-E/k_B T)$ with the density of states $N(E)$ and the partition function $Q$.

In the low pressure limit, the rate is given by a second order expression in [A] and [M], and thus from Eq.(6) we have the unimolecular reaction rate coefficient at the energy $E$ [3],

$$k_{uni}^{0}(E) = \lim_{[M] \to 0} \frac{\upsilon}{[A][M]} = dk_1(E) , \qquad (10)$$

and the thermal rate coefficient is $k_{uni}^{0} = k_1$. If the statistical property of the system is assumed to follow Boltzmann–Gibbs (BG) distribution [3], it is

$$k_{uni}^{0} = k_1 = \frac{Z_{AM}}{Q} \int_{E_0}^{\infty} N(E) \exp\left(-\frac{E_0}{k_B T}\right) dE , \qquad (11)$$

where $Z_{AM}$ is the collision number between A and M per unit volume per unit time, $Q$ is the partition function, $N(E)$ is the density of states of the reactant, $k_B$ is the Boltzmann constant, and $T$ is temperature.

There are many reaction rate theories that have been studied to obtain possible unimolecular reaction rate coefficients in the cases of both high pressure limit and low pressure limit. In all the traditional theories, just as that in Eq. (11), the energy distribution function $P(E)$ in Eq. (9) has always been assumed to be a BG distribution. These are only a good approximation in the situations when the statistical property of the system can be described by BG statistical mechanics. However, chemical reaction systems are generally far from thermal equilibrium and therefore the statistical properties do not exactly obey BG distribution. In fact, the non-BG distributions and/or power-law distributions in chemical reacting systems have been noted and studied both experimentally and theoretically in the processes such as fluctuations within a single protein molecule [4], the single-molecule conformation dynamics [5-7], single-molecule fluorescence intermittency [8], reaction-diffusion processes [9], chemical reactions [10], combustion processes [11], gene expressions [12], cell reproductions [13], complex cellular networks [14], and small organic molecules [15]. In these processes, the reaction rate coefficients may be energy-dependent (and/or time-dependent [16-18]) power-law forms [19, 20]. In these situations, the conventional reaction rate coefficients should be modified on the basis of the corresponding non-BG energy distributions.

On the other hand, the statistical mechanical theory of power-law distributions has been developed. For example, the generalized Gibbsian theory for power-law distributions was presented to the systems away from equilibrium [21]. In stochastic dynamical theory on Brownian motion in a complex system, power-law distributions can also be discovered by introducing generalized fluctuation-dissipation relations and solving Fokker-Planck equations [22, 23]. It is especially worth mentioning that, in recent years, nonextensive statistical mechanics (NSM) based on Tsallis entropy has received great attention and very wide applications in a variety of interesting problems in physics, chemistry, astronomy, biology, engineering and technology etc [24]. In NSM, the power-law distribution can be derived using the extremization of Tsallis entropy. When one generalizes BG statistical mechanics to NSM, the usual exponential and logarithm can be replaced respectively by the $q$-exponential and the $q$-logarithm [24]. For instance, the $q$-exponential can be defined as

$$\exp_q x = \left[1 + (q-1)x\right]^{1/(q-1)}, \quad (\exp_1 x = e^x),$$

if $1 + (q-1)x > 0$ and as $\exp_v x = 0$ otherwise. And the inverse function, the $q$-logarithm can be defined as

$$\ln_q x = \frac{x^{q-1} - 1}{q - 1}, \quad (x > 0, \ \ln_1 = \ln x).$$

NSM has been studied as a reasonable generalization of BG statistical mechanics, which has been able to produce a statistical description of a nonequilibrium stationary state in the interacting systems, and so becomes a very useful tool to approach complex systems whose properties go beyond the realm governed by BG statistical mechanics. Most recently, a transition state theory (TST) for the systems with power-law distributions was studied and the generalized reaction rate formulae were presented for one-dimensional and $n$-dimensional Hamiltonian systems away from equilibrium [19]. And the power-law TST reaction rate coefficient for an elementary bimolecular reaction [25] and the collision theory reaction rate coefficient for power-law distributions [26] were also studied if the reaction takes place in a nonequilibrium system with power-law distributions. In addition, the mean first passage time [27], and the escape rate for power–law distributions in both overdamped systems and low–to–intermediate damping [28, 29] were studied. These developments now naturally give rise to a possibility to generalize the unimolecular reaction rate theories to the nonequilibrium systems with power-law distributions.

The purpose of this work is to generalize the rate coefficients (9) and (11) to the systems with power-law distributions. The paper is organized as follows. In Section 2, we study the power-law unimolecular reaction rate coefficient. In Section 3, we make numerical analyses to show the dependence of the power-law reaction rate coefficient on the quantities such as power-law parameter, and temperature. As application examples of the new theory, in Section 4 we calculate the power-law reaction rate coefficients of several reactions, compare them with the values in experiment studies, and determine the power-law parameter. Finally, in Section 5 we give the conclusion and discussion.

## 2. The unimolecular reaction rate theory for power-law distributions

In the high pressure limit the key problem is how to calculate the rate coefficient $k(E)$ in the unimolecular decay reaction (3). There are a lot of theories dealing with this problem; among them the most common one was developed by Rice, Ramsperger, Kassel, and Marcus, known as RRKM theory. In this theory, one focuses on the transition state [2],

$$A^* \rightarrow A^{\neq} \rightarrow \text{Products}, \qquad (12)$$

where $A^{\neq}$ is transition state (also called activated complex).

RRKM theory basically applies transition state theory to a unimolecular reaction. The essential assumptions in the theory are equivalent to those in the conventional transition state theory: (a) The reaction coordinate can be separated from the other degrees of freedom in the saddle point region, and the motion in the reaction coordinate can be described by classical mechanics. (b) A point of no return (corresponding to the transition state) exists along the reaction coordinate, i.e., there are no recrossings of trajectories at this point. (c) The energy states of the reactants as well as the transition state are populated according to the BG distribution. Besides the three assumptions, there is a key assumption in RRKM theory that all possible ways of partitioning a given total energy between the internal degrees of freedom of the transition state and the translational energy of the reaction coordinate are equally probable [2]. The expression for the rate coefficient in RRKM theory [2, 3] is

$$k(E) = \frac{G^{\neq}(E - E_0)}{hN(E)}, \qquad (13)$$

where $G^{\neq}(E - E_0)$ is the sum of states in the transition state at and below the energy $E - E_0$ that the reaction coordinate is not taken into account, $N(E)$ is the density of states of the reactant, and $h$ is Planck's constant. For a reaction system with $s$ degrees of freedom, the sum of states of $s$ uncoupled harmonic oscillators is

$$G(E) = \frac{E^s}{s! \prod_{i=1}^{s} h\nu_i}.$$

Then we have

$$G^{\neq}(E - E_0) = \frac{(E - E_0)^{s-1}}{(s-1)! \prod_{i=1}^{s-1} h\nu_i^{\neq}}, \qquad$$

(14)

and the density of states of $s$ uncoupled harmonic oscillators,

$$N(E) = \frac{dG(E)}{dE} = \frac{E^{s-1}}{(s-1)! \prod_{i=1}^{s} h\nu_i}, \qquad (15)$$

Now, if there are $s$ degrees of freedom in the reactant with frequencies $\nu_i$, there are $s-1$ degrees of freedom in the transition state with frequencies $\nu_i^{\neq}$ when the reaction coordinate is excluded. Substituting (14) and (15) into (13), we get

$$k(E) = \left(\frac{E - E_0}{E}\right)^{s-1} \frac{\prod_{i=1}^{s} \nu_i}{\prod_{i=1}^{s-1} \nu_i^{\neq}}. \qquad (16)$$

Based on NSM, the distribution function in Eq.(9) for the system with power-law distribution can be generalized as

$$P(E) = \frac{N(E)}{Q} \left[1 - (\nu - 1)\frac{E}{k_B T}\right]_+^{1/(\nu-1)}, \qquad (17)$$

where the partition function becomes

$$Q = \int_0^\infty N(E) \left[ 1 - (\nu - 1) \frac{E}{k_B T} \right]_+^{1/(\nu-1)} dE \,, \tag{18}$$

and $[y]_+ = y$ for y > 0, and zero otherwise. Integrating Eq. (18) we have,

$$Q = \frac{(k_B T)^s}{\prod_{i=1}^s h\nu_i} \begin{cases} \dfrac{\Gamma\left(\dfrac{1}{1-\nu} - s\right)}{(1-\nu)^s \, \Gamma\left(\dfrac{1}{1-\nu}\right)}, & 1 - \dfrac{1}{s} < \nu < 1, \\[4mm] \dfrac{\Gamma\left(\dfrac{1}{\nu-1} + 1\right)}{(\nu-1)^s \, \Gamma\left(\dfrac{1}{\nu-1} + 1 + s\right)}, & \nu > 1. \end{cases} \tag{19}$$

when we take $\nu \to 1$, Eq. (17) is a BG distribution and the partition function becomes $Q = (k_B T)^s / \prod_{i=1}^s h\nu_i$. The power-law $\nu$-distribution represents the statistical property of a nonequilibrium system being at a stationary-state and the parameter $\nu \neq 1$ measures a distance away from equilibrium [22].

Substituting Eq. (15) into Eq. (17) we get

$$P(E) = \frac{E^{s-1}}{Q(s-1)! \prod_{i=1}^s h\nu_i} \left[ 1 - (\nu - 1) \frac{E}{k_B T} \right]_+^{1/(\nu-1)} . \tag{20}$$

Substituting Eq.(16) and Eq.(20) into Eq.(9), we obtain the generalized unimolecular reaction rate coefficient in the high pressure limit for the system with the power-law distribution,

$$k_{\text{uni}}^\infty = \frac{\prod_{i=1}^s \nu_i}{\prod_{i=1}^{s-1} \nu_i^{\neq}} \left[ 1 - (\nu - 1) \frac{E_0}{k_B T} \right]_+^{1/(\nu-1)+s} . \tag{21}$$

It is clear that by taking the limit $\nu \to 1$ in Eq. (21) the standard RRKM rate coefficient for the system with a BG distribution [3] can be recovered perfectly,

$$k_{BG}^\infty = \frac{\prod_{i=1}^s \nu_i}{\prod_{i=1}^{s-1} \nu_i^{\neq}} \exp\left( -\frac{E_0}{k_B T} \right) . \tag{22}$$

In the same way, in the low pressure limit, the unimolecular reaction rate coefficient (11) for the system with power-law distributions can be generalized as

$$k_{\text{uni}}^0 = \frac{Z_{\text{AM}}}{Q} \int_{E_0}^\infty N(E) \left[ 1 - (\nu - 1) \frac{E_0}{k_B T} \right]_+^{1/(\nu-1)} dE \,, \tag{23}$$

where the partition function $Q$ is the same as Eq.(18). By calculation we finally obtain the generalized rate coefficient,

$$k_{\text{uni}}^0 = \frac{Z_{\text{AM}}}{(s-1)!} \begin{cases} B\left[\frac{(1-\nu)E_0}{k_B T + (1-\nu)E_0}, s, \frac{1}{1-\nu} - s\right] \dfrac{\Gamma\left(\dfrac{1}{1-\nu}\right)}{\Gamma\left(\dfrac{1}{1-\nu} - s\right)}, & 1 - \frac{1}{s} < \nu < 1, \\[4mm] B\left[\frac{(\nu-1)E_0}{k_B T}, s, \frac{1}{\nu-1} + 1\right] \dfrac{\Gamma\left(\dfrac{1}{\nu-1} + 1 + s\right)}{\Gamma\left(\dfrac{1}{\nu-1} + 1\right)}, & \nu > 1, \end{cases}$$
(24)

where $B$ is the upper-part incomplete Beta function. When we take the limit $\nu \to 1$ in Eq.(24) (see the Appendix), we can recover the rate coefficient for the system with a BG distribution [30],

$$k_{BG}^0 = \frac{Z_{\text{AM}}}{(s-1)!} \Gamma\left(\frac{E_0}{k_B T}, s\right),$$
(25)

where $\Gamma(E_0/k_B T, s)$ is the upper-part incomplete Gamma function [31],

$$\Gamma(x_0, s) = \int_{x_0}^{\infty} x^{s-1} \exp(-x)\, dx.$$

If $E_0 \gg k_B T$, i.e., $x_0 \gg 1$, using $\int_{x_0}^{\infty} x^n \exp(-x)\, dx = x_0^n \exp(-x_0)$, Eq.(25) becomes the familiar form of the rate coefficient in the conventional unimolecular rate theory [3, 32],

$$k_{BG}^0 = \frac{Z_{\text{AM}}}{(s-1)!} \left(\frac{E_0}{k_B T}\right)^{s-1} \exp\left(-\frac{E_0}{k_B T}\right).$$
(26)

## 3. Numerical analyses of the power-law rate coefficient

In order to illustrate the dependence of the power-law unimolecular reaction rate coefficients on the physical quantities in the high and low pressure limits, such as the power-law $\nu$–parameter, the threshold energy $E_0$, the temperature $T$, and the number of degrees of freedom $s$, we have made numerical analyses of the unimolecular reaction rate coefficients with regard to $\nu$, $E_0$, $T$, and $s$. The generalized rate coefficients, Eq. (21) and Eq. (24), are shown in Figs 1-8 as a function of these quantities, respectively. In these numerical analyses, when one of these quantities was chosen as a variable, the other quantities were fixed. The fixed data were taken as typical data in the chemical reactions. The fixed values were chosen as $E_0$=50kJ/mol, $T$=300K and $s = 10$.

Fig.1 and Fig.2 illustrated the dependence of the generalized rate coefficients of the unimocular reaction on the parameter $\nu$ in high and low pressure limits, respectively. The range of the $\nu$-axis was chosen near the unity, implying a state not very far away from equilibrium. The numerical analyses showed a very strong dependence of the rate coefficients on the parameter $\nu$, which imply that a tiny deviation from a BG distribution and thus from thermal equilibrium would result in a significant variation in the rate coefficient. Such high sensitivity of the reaction rate to the $\nu$-parameter has shown the important role of the power-law distribution in the calculation of reaction rate coefficient.

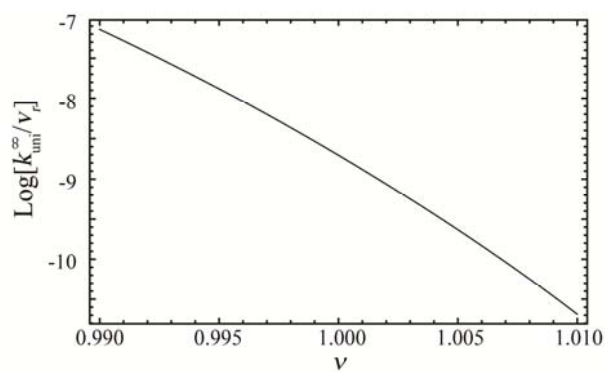

Fig.1. In the high pressure limit, the rate coefficient is as a function of $\nu$

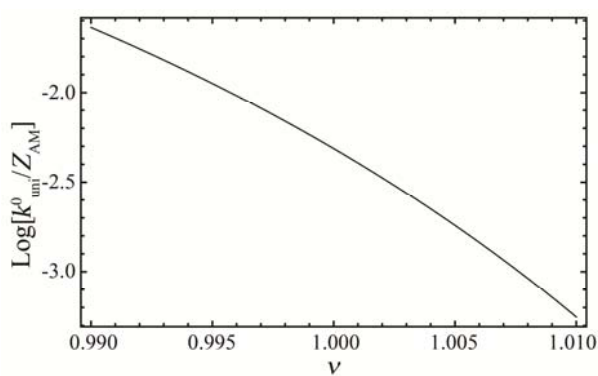

Fig.2. In the low pressure limit, the rate coefficient is as a function of $\nu$

Fig.3 and Fig.4 illustrated the dependence of the generalized rate coefficients of the unimocular reaction on the threshold energy $E_0$. The range of $E_0$-axis was chosen 10~100kJ/mol, as the typical threshold energy in the chemical reactions. The $\nu$–parameter was taken with three different values. The rate coefficient for $\nu$=1 corresponds to that for the system with a BG distribution. In the high pressure limit, the line for $\nu = 1$ is exactly a straight line, corresponding to the rate of the famous Arrhenius law.

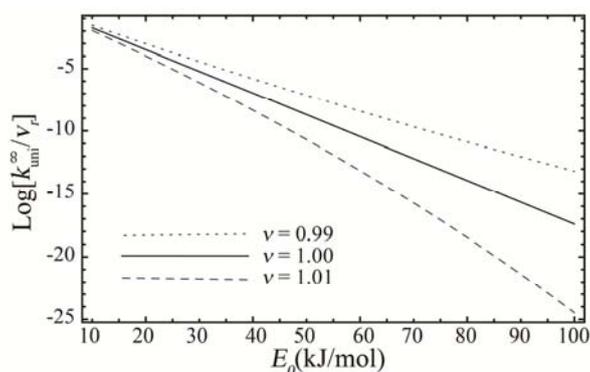

Fig.3. In the high pressure limit, the rate coefficient is as a function of $E_0$
for three values of $\nu$

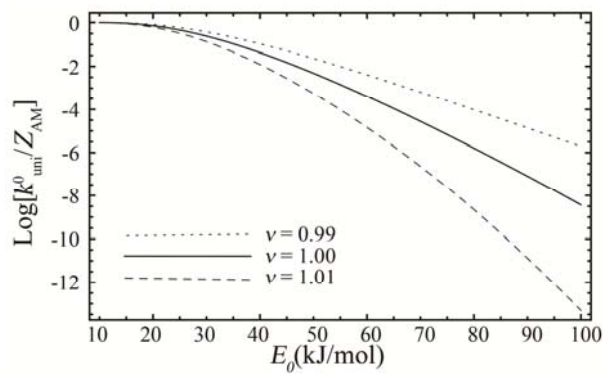

Fig.4. In the low pressure limit, the rate coefficient is as a function of $E_0$ for three values of $\nu$

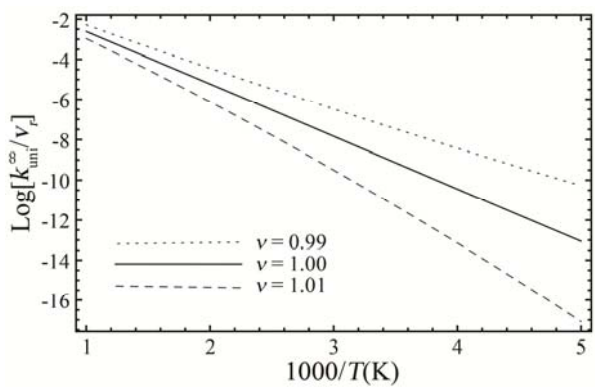

Fig.5. In the high pressure limit, the rate coefficient is as a function of $T$ for three values of $\nu$

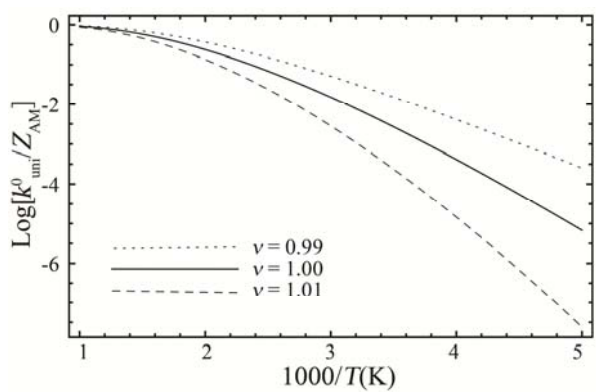

Fig.6. In the low pressure limit, the rate coefficient is as a function of $T$ for three values of $\nu$

Fig.5 and Fig.6 illustrated the dependence of the generalized rate coefficients of the unimolecular reaction on the temperature $T$ for three values of $\nu$. The range of $T$-axis was chosen 200~1000K, as the typical temperature range in the chemical reactions. The rate coefficient for $\nu=1$ corresponds to that for the system with a BG distribution. Also in the high pressure limit, the line for $\nu=1$ is exactly a straight line, corresponding to the rate of the famous Arrhenius law.

Fig.7 and Fig.8 illustrated the dependence of the generalized rate coefficients of the unimolecular reaction on the number of degrees of freedom $s$ for three values of $\nu$. The range of $s$-axis was chosen 1~30, as the typical number of degrees of freedom range in molecules. It is shown that in Fig.7, the dependences of the rate coefficient on $s$ are different evidently for the case $\nu > 1$ and the case $\nu < 1$. For the case $\nu > 1$, the rate coefficient decreases gradually as $s$ increases, but for the case $\nu < 1$, it increases gradually as $s$ increases. The lines for $\nu=1$ correspond to the conventional unimolecular reaction rate coefficient in BG statistical mechanics.

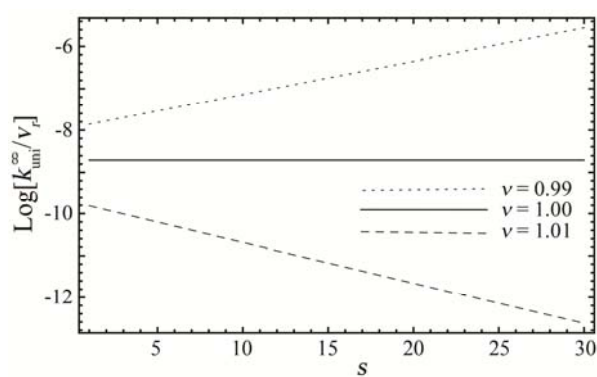

Fig.7. In the high pressure limit, the rate coefficient is as a function of $s$
for three values of $\nu$

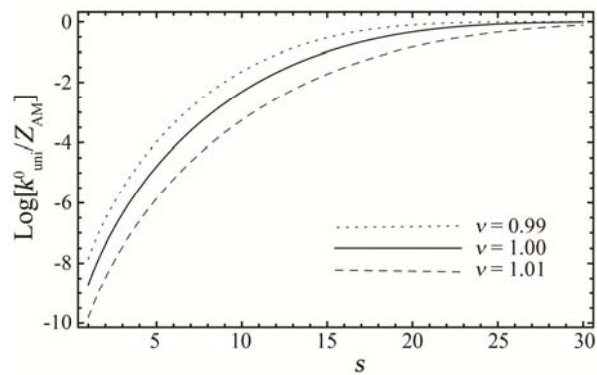

Fig.8. In the low pressure limit, the rate coefficient is as a function of $s$
for three values of $\nu$

## 4. The application to two examples of unimolecular reactions.

We have generalized the unimolecular reaction rate theory to the systems with the power-law distributions described in NSM, and have derived the power-law rate coefficients, Eq. (21) in high pressure limit and Eq. (24) in low pressure limit, respectively. Now we can take two unimolecular reactions, $CH_3CO \rightarrow CH_3+CO$ and CH₃NC→CH₃CN, as application examples to calculate the power-law rate coefficients. Usually, because the collisions with impurities or the collisions between reactant molecules may become important, experimentally the low pressure limit is difficult to attain [3]. We here only give a test of the power-law rate coefficient in the high pressure limit, Eq. (21), where the vibrational frequencies of the reactants and the transition state of the reactions were listed in Table 1 and Table 2, respectively. Note that the imaginary frequency of the transition state is not included in the tables.

Table 1. The vibrational frequencies for the reactant and transition state
of the reaction $CH_3CO \rightarrow CH_3 + CO$ [2]

| $v_{CH3CO}$/cm⁻¹ | 3193 | 3188 | 3175 | 1928 | 1478 | 1477 | 1361 | 1062 | 960 | 884 | 468 | 101 |
|---|---|---|---|---|---|---|---|---|---|---|---|---|
| $v_{TS}$/ cm⁻¹ | 3325 | 3321 | 3225 | 2027 | 1452 | 1445 | 901 | 568 | 566 | 276 | 43 | |

Table 2. The vibrational frequencies for the reactant and transition state
of the reaction $CH_3NC \rightarrow CH_3CN$ [30]

| $v_{CH3NC}$/cm⁻¹ | 3014 | 3014 | 2966 | 2166 | 1467 | 1429 | 1129 | 945 | 263 | 263 |
|---|---|---|---|---|---|---|---|---|---|---|
| $v_{TS}$/ cm⁻¹ | 3070 | 3049 | 2945 | 1970 | 1439 | 1311 | 975 | 956 | 609 | |

The experimental measurements of the rate coefficients for these two reactions can be found in the NIST chemical kinetics database at http://kinetics.nist.gov/kinetics. As stated previously, in the high pressure limit, the rate is given by the first order expression in [A]. Thus we can only take the first order expression of the experimental data in NIST. The temperature range, according to the experimental data in NIST, for the reaction CH₃CO→CH₃+CO is 296-568K, and for the reaction CH₃NC→CH₃CN is 393-593K.

Table 3. The experimental and the calculated values of the rate coefficient
of the reaction $CH_3CO \rightarrow CH_3 + CO$

| $T$(K) | $k_{BG}^{\infty}$ (s⁻¹) | $k_{exp}$(s⁻¹) | $\delta$ | $k_{uni}^{\infty}$ (s⁻¹) | $v$ |
|---|---|---|---|---|---|
| 296 | 7.2 | 1.5 | 380% | 1.5 | 1.00173 |
| 400 | 2.8×10⁴ | 1.1×10⁴ | 155% | 1.1×10⁴ | 1.00167 |
| 500 | 3.1×10⁶ | 1.7×10⁶ | 82% | 1.7×10⁶ | 1.00146 |
| 568 | 2.9×10⁷ | 1.9×10⁷ | 53% | 1.9×10⁷ | 1.00126 |

Table 4. The experimental and the calculated values of the rate coefficient
of the reaction $CH_3NC \rightarrow CH_3CN$

| $T$(K) | $k_{BG}^{\infty}$ (s⁻¹) | $k_{exp}$(s⁻¹) | $\delta$ | $k_{uni}^{\infty}$ (s⁻¹) | $v$ |
|---|---|---|---|---|---|
| 393 | 6.3×10⁻⁹ | 1.6×10⁻⁸ | 61% | 1.6×10⁻⁸ | 0.9994 |
| 500 | 1.8×10⁻⁴ | 5.5×10⁻⁴ | 67% | 5.5×10⁻⁴ | 0.9990 |
| 593 | 0.069 | 0.23 | 70% | 0.23 | 0.9985 |

In Table 3 and Table 4, we listed the experimental values $k_{\exp}$ and the calculation values of the rate coefficients for these two reactions, where $k_{BG}^{\infty}$ is the rate coefficient calculated using the old formula Eq.(22) in BG statistical mechanics, $k_{uni}^{\infty}$ is the rate coefficient calculated using the new formula Eq.(21) in NSM, the quantity $\delta$ is the relative error of $k_{BG}^{\infty}$ to $k_{\exp}$, defined by $\delta = |k_{BG}^{\infty} - k_{\exp}| / k_{\exp}$, and $\nu$ is the fitted power-law parameter.

We find that there were large relative errors $\delta$ of $k_{BG}^{\infty}$ to $k_{\exp}$ using the old formula Eq.(22), but $k_{uni}^{\infty}$ using the new formula Eq.(21) with the $\nu$-parameter slightly different from unity could be exactly in agreement with all the experimental studies in the temperature range. The $\nu$-parameter varies as the temperature varies and there is one $\nu$-parameter in the experimental measurements at one temperature. Such a variation in the $\nu$-parameter is a result of the fact that the $\nu$-parameter depends on the intermolecular interactions and the temperature distribution in the system, mirroring the differences between the experiments at different temperatures and the environment. According to Du's equation [33, 34], the power-law parameter different from one is related to the temperature and interaction potential gradient in a nonequilibrium system.

## 5. Conclusions and discussions

The unimolecular reaction theory for the systems with power-law distributions is beyond the scope of conventional unimolecular reaction theories such as RRKM for the systems with a BG distribution, and therefore if chemical reactions occur in the systems with power-law distributions the conventional rate formulae need to be modified. In conclusion, we have studied the unimolecular reaction rate coefficients of the reactions taking place in a nonequilibrium system with the power-law distribution, generalized the unimolecular reaction rate theory to the systems with the power-law distributions based on NSM, and derived the power-law rate coefficients, Eq. (21) in the high pressure limit and Eq.(24) in the low pressure limit, respectively. New rate coefficient formulae depend on the power-law parameter $\nu \neq 1$, and when $\nu \rightarrow 1$ restore the conventional rate formulae.

We have made numerical analyses to illustrate the dependence of the power-law rate coefficients of unimolecular reactions on the relevant physical quantities, such as the power-law parameter $\nu$, the threshold energy $E_0$, the temperature $T$, and the number of degrees of freedom, in the high pressure limit and in the low pressure limit, respectively. We have clearly shown a very strong dependence of the rate coefficients on the power-law parameter, and indicated that a tiny deviation from the BG distribution (thus from thermodynamic equilibrium) would result in a significant effect on the reaction rate. Such high sensitivity of the reaction rate coefficient to the power-law parameter showed the important role of the power-law distributions in calculations of the reaction rate coefficients of unimolecular reactions.

In order to show the application of the new rate formulae to the unimolecular reactions taking place in the system with the power-law distributions, we have taken two unimolecular reactions, $CH_3CO \rightarrow CH_3+CO$ and $CH_3NC \rightarrow CH_3CN$, as application examples to calculate their power-law rate coefficients. We obtained the rate coefficients with the $\nu$-parameter slightly different from one, which could be exactly in agreement with all the experimental studies on these two reactions in the given temperature ranges.

### Appendix

In this appendix we give the details of Eq. (25) derived from Eq. (24) when $\nu \rightarrow 1$. The lower-part incomplete Beta function $B'$ and Gamma function $\Gamma'$, and the relation between the upper-part and lower-part function [31] are used,

$$B(a,b,c) + B'(a,b,c) = B(b,c), \tag{A.1}$$

and

$$\Gamma(a,b) + \Gamma'(a,b) = \Gamma(b). \tag{A.2}$$

A relation between the lower-part incomplete Beta function and the lower-part incomplete Gamma function is defined [31] as

$$B'(a,b,c)\Gamma'(d,b+c) = \int_0^a x^{b-1}(1-x)^{c-1}dx \int_0^d t^{b+c-1}\exp(-t)dt, \tag{A.3}$$

where $0 < a < 1$ and $d$ is positive. Using the replacement $u = xt$ and $v = t(1-x)$, (A.3) becomes

$$B'(a,b,c)\Gamma'(d,b+c) = \int_0^{ad}du \int_0^d dv \exp(-u-v)u^{b-1}v^{c-1}$$

$$= \int_0^{ad}\exp(-u)u^{b-1}du \int_0^d \exp(-v)v^{c-1}dv$$

$$= \Gamma'(ad,b)\Gamma'(d,c), \tag{A.4}$$

and then we have

$$B'(a,b,c) = \frac{\Gamma'(ad,b)\Gamma'(d,c)}{\Gamma'(d,b+c)}. \tag{A.5}$$

Another relation between complete Beta function and gamma function [31] is

$$B(b,c) = \frac{\Gamma(b)\Gamma(c)}{\Gamma(b+c)}. \tag{A.6}$$

Using Eq.(A.1), Eq.(A.5) and Eq.(A.6), we can write Eq.(24) as

$$k_{\text{uni}}^0(T) = \frac{Z_{\text{AM}}}{(s-1)!}\begin{cases} \Gamma(s) - \Gamma'\left(\frac{E_0}{k_B T},s\right)\dfrac{\Gamma'\left(\dfrac{1}{1-\nu}+\dfrac{E_0}{k_B T},\dfrac{1}{1-\nu}-s\right)\Gamma\left(\dfrac{1}{1-\nu}\right)}{\Gamma'\left(\dfrac{1}{1-\nu}+\dfrac{E_0}{k_B T},\dfrac{1}{1-\nu}\right)\Gamma\left(\dfrac{1}{1-\nu}-s\right)}, 1-\dfrac{1}{s}<\nu<1, \\[3em] \Gamma(s) - \Gamma'\left(\frac{E_0}{k_B T},s\right)\dfrac{\Gamma'\left(\dfrac{1}{\nu-1},\dfrac{1}{\nu-1}+1\right)\Gamma\left(\dfrac{1}{\nu-1}+1+s\right)}{\Gamma'\left(\dfrac{1}{\nu-1},\dfrac{1}{\nu-1}+1+s\right)\Gamma\left(\dfrac{1}{\nu-1}+1\right)}, \quad \nu>1. \end{cases}$$

For the lower-part incomplete gamma function, repeated application of the recurrence relation leads to the power series expansion [35],

$$\Gamma'(a,b) = a^b \Gamma(b) e^{-a} \sum_{k=0}^{\infty} \frac{a^k}{\Gamma(b+k+1)} .$$ (A.8)

By using Eq. (A.8) in Eq. (A.7), we have

$$\frac{\Gamma'\left(\frac{1}{1-\nu}+\frac{E_0}{k_BT}, \frac{1}{1-\nu}-s\right)\Gamma\left(\frac{1}{1-\nu}\right)}{\Gamma'\left(\frac{1}{1-\nu}+\frac{E_0}{k_BT}, \frac{1}{1-\nu}\right)\Gamma\left(\frac{1}{1-\nu}-s\right)}$$

$$=\sum_{k=0}^{\infty}\frac{\left(\frac{1}{1-\nu}+\frac{E_0}{k_BT}\right)^{k-s}}{\Gamma\left(\frac{2-\nu}{1-\nu}-s+k\right)} \bigg/ \sum_{k=0}^{\infty}\frac{\left(\frac{1}{1-\nu}+\frac{E_0}{k_BT}\right)^{k}}{\Gamma\left(\frac{2-\nu}{1-\nu}+k\right)} , \quad \text{for } 1-\frac{1}{s}<\nu<1,$$ (A.9)

and

$$\frac{\Gamma'\left(\frac{1}{\nu-1}, \frac{1}{\nu-1}+1\right)\Gamma\left(\frac{1}{\nu-1}+1+s\right)}{\Gamma'\left(\frac{1}{\nu-1}, \frac{1}{\nu-1}+1+s\right)\Gamma\left(\frac{1}{\nu-1}+1\right)}$$

$$=\sum_{k=0}^{\infty}\frac{\left(\frac{1}{\nu-1}\right)^{k-s}}{\Gamma\left(\frac{2\nu-1}{\nu-1}+k\right)} \bigg/ \sum_{k=0}^{\infty}\frac{\left(\frac{1}{\nu-1}\right)^{k}}{\Gamma\left(\frac{2\nu-1}{\nu-1}+s+k\right)} , \quad \text{for } \nu>1.$$ (A.10)

If we let $k-s=t$ in the numerators of Eqs.(A.9) and (A.10), they can be rewritten respectively as

$$(\sum_{t=-s}^{-1}+\sum_{t=0}^{\infty})\frac{\left(\frac{1}{1-\nu}+\frac{E_0}{k_BT}\right)^{t}}{\Gamma\left(\frac{2-\nu}{1-\nu}+t\right)} \bigg/ \sum_{k=0}^{\infty}\frac{\left(\frac{1}{1-\nu}+\frac{E_0}{k_BT}\right)^{k}}{\Gamma\left(\frac{2-\nu}{1-\nu}+k\right)} , \text{for } 1-\frac{1}{s}<\nu<1,$$ (A.11)

and

$$(\sum_{t=-s}^{-1}+\sum_{t=0}^{\infty})\frac{\left(\frac{1}{\nu-1}\right)^{t}}{\Gamma\left(\frac{2\nu-1}{\nu-1}+s+t\right)} \bigg/ \sum_{k=0}^{\infty}\frac{\left(\frac{1}{\nu-1}\right)^{k}}{\Gamma\left(\frac{2\nu-1}{\nu-1}+s+k\right)} , \quad \text{for } \nu>1.$$ (A.12)

When $\nu \to 1$, Eqs. (A.11) and (A.12) both converge to 1, and then (A.7) becomes

$$k_{\text{uni}}^{0}(T) = \frac{Z_{\text{AM}}}{(s-1)!}\left[\Gamma(s) - \Gamma'\left(\frac{E_0}{k_BT}, s\right)\right].$$ (A.13)

Using (A.2) we get Eq. (25).

**Acknowledgment**

This work is supported by the National Natural Science Foundation of China under Grant No. 11175128 and also by the Higher School Specialized Research Fund for Doctoral Program under Grant No. 20110032110058.